\documentclass[graybox]{svmult}

\usepackage{mathptmx}       
\usepackage{helvet}         
\usepackage{courier}        
\usepackage{type1cm}        
\usepackage{makeidx}         
\usepackage{natbib}
\usepackage{graphicx}        
\usepackage{multicol}        
\usepackage[bottom]{footmisc}

\makeindex             

\usepackage{bm}

\newcommand{\EQ}{\begin{equation}}
\newcommand{\EN}{\end{equation}}
\newcommand{\EQA}{\begin{eqnarray}}
\newcommand{\ENA}{\end{eqnarray}}
\newcommand{\eq}[1]{(\ref{#1})}

\newcommand{\Eq}[1]{Equation~(\ref{#1})}
\newcommand{\Eqs}[2]{Equations~(\ref{#1}) and~(\ref{#2})}

\newcommand{\Sec}[1]{Section~\ref{#1}}

\newcommand{\Fig}[1]{Figure~\ref{#1}}
\newcommand{\FFig}[1]{Figure~\ref{#1}}

\newcommand{\bra}[1]{\langle #1\rangle}

\newcommand{\meanrho}{\overline{\rho}}

\newcommand{\meanFFFF}{\overline{\mbox{\boldmath ${\cal F}$}}{}}{}

\newcommand{\meanEMF}{\overline{\bf{\cal{E}}}}
\newcommand{\meanFF}{\overline{\boldmath{\mathcal{F}}}}
\newcommand{\meanFFm}{\overline{\boldmath{\mathcal{F}}}_{\rm m}}
\newcommand{\meanFFf}{\overline{\boldmath{\mathcal{F}}}_{\rm f}}
\newcommand{\meanAA}{\overline{\bm{A}}}
\newcommand{\meanBB}{\overline{\bm{B}}}
\newcommand{\meanJJ}{\overline{\bm{J}}}
\newcommand{\meanUU}{\overline{\bm{U}}}

\newcommand{\meanB}{\overline{B}}

\newcommand{\meanh}{\overline{h}}

\newcommand{\meanhf}{\overline{h}_{\rm f}}

{}

\newcommand{\alphaK}{\alpha_{\rm K}}
\newcommand{\alphaM}{\alpha_{\rm M}}

\newcommand{\nullvector}{{\bf0}}
\newcommand{\nab}{\bm{\nabla}}

\newcommand{\AAA}{\bm{A}}
\newcommand{\BB}{\bm{B}}
\newcommand{\JJ}{\bm{J}}
\newcommand{\SSS}{\bm{S}}
\newcommand{\UU}{\bm{U}}
\newcommand{\aaaa}{\bm{a}}
\newcommand{\bb}{\bm{b}}
\newcommand{\ee}{\bm{e}}
\newcommand{\jj}{\bm{j}}
\newcommand{\uu}{\bm{u}}
\newcommand{\vv}{\bm{v}}
\newcommand{\oo}{\bm{\omega}}

\newcommand{\ii}{{\rm i}}

\newcommand{\dive}{{\rm div}  \, {}}

\newcommand{\dd}{{\rm d} {}}

\newcommand{\const}{{\rm const}  {}}

\def\la{\mathrel{\mathchoice {\vcenter{\offinterlineskip\halign{\hfil
$\displaystyle##$\hfil\cr<\cr\sim\cr}}}
{\vcenter{\offinterlineskip\halign{\hfil$\textstyle##$\hfil\cr<\cr\sim\cr}}}
{\vcenter{\offinterlineskip\halign{\hfil$\scriptstyle##$\hfil\cr<\cr\sim\cr}}}
{\vcenter{\offinterlineskip\halign{\hfil$\scriptscriptstyle##$\hfil\cr<\cr\sim\cr}}}}}
\def\ga{\mathrel{\mathchoice {\vcenter{\offinterlineskip\halign{\hfil
$\displaystyle##$\hfil\cr>\cr\sim\cr}}}
{\vcenter{\offinterlineskip\halign{\hfil$\textstyle##$\hfil\cr>\cr\sim\cr}}}
{\vcenter{\offinterlineskip\halign{\hfil$\scriptstyle##$\hfil\cr>\cr\sim\cr}}}
{\vcenter{\offinterlineskip\halign{\hfil$\scriptscriptstyle##$\hfil\cr>\cr\sim\cr}}}}}
\def\Atan{\mbox{\rm Arctan}}

\def\Pm{\mbox{\rm Pr}_M}
\def\Rm{\mbox{\rm Re}_M}
\def\RM{{\rm RM}}

\def\EM{E_{\rm M}}

\def\kf{k_{\rm f}}
\def\HM{H_{\rm M}}
\def\EM{E_{\rm M}}
\def\epsf{\epsilon_{\rm f}}

\def\kM{k_{\rm M}}

\def\urms{u_{\rm rms}}

\def\kappah{\kappa_{\rm h}}

\def\etat{\eta_{\rm t}}

\def\half{{\textstyle{1\over2}}}

\newcommand{\G}{\,{\rm G}}

\newcommand{\g}{\,{\rm g}}
\newcommand{\s}{\,{\rm s}}

\newcommand{\cm}{\,{\rm cm}}

\newcommand{\km}{\,{\rm km}}

\newcommand{\kms}{\,{\rm km/s}}

\newcommand{\pc}{\,{\rm pc}}
\newcommand{\kpc}{\,{\rm kpc}}

\newcommand{\Myr}{\,{\rm Myr}}

\newcommand{\erg}{\,{\rm erg}}

\newcommand{\yan}[5]{, #5, {Astron.\ Nachr.\ }{\bf #2}, #3-#4 (#1)}

\newcommand{\yana}[5]{, #5, {Astron.\ Astrophys.\ }{\bf #2}, #3-#4 (#1)}
\newcommand{\yanaN}[4]{, #4, {Astron.\ Astrophys.\ }{\bf #2}, #3 (#1)}

\newcommand{\ysci}[5]{, #5, {Science }{\bf #2}, #3-#4 (#1)}

\newcommand{\ysov}[5]{, #5, {Sov.\ Astron.\ }{\bf #2}, #3-#4 (#1)}

\newcommand{\ymn}[5]{, #5, {Monthly Notices Roy.\ Astron.\ Soc.\ }{\bf #2}, #3-#4 (#1)}

\newcommand{\pmn}[3]{, #2, {Monthly Notices Roy.\ Astron.\ Soc.}, in press, {arXiv:#3} (#1)}

\newcommand{\yqjras}[5]{, #5, {Quart.\ J.\ Roy.\ Astron.\ Soc.\ }{\bf #2}, #3-#4 (#1)}

\newcommand{\yjfm}[5]{, #5, {J.\ Fluid Mech.\ }{\bf #2}, #3-#4 (#1)}

\newcommand{\yprd}[5]{, #5, {Phys.\ Rev.\ D }{\bf #2}, #3-#4 (#1)}
\newcommand{\ypre}[5]{, #5, {Phys.\ Rev.\ E }{\bf #2}, #3-#4 (#1)}
\newcommand{\yprdN}[4]{, #4, {Phys.\ Rev.\ D }{\bf #2}, #3 (#1)}
\newcommand{\ypreN}[4]{, #4, {Phys.\ Rev.\ E }{\bf #2}, #3 (#1)}
\newcommand{\ypreNS}[4]{, #4 {Phys.\ Rev.\ E }{\bf #2}, #3 (#1)}
\newcommand{\yprl}[5]{, #5, {Phys.\ Rev.\ Lett.\ }{\bf #2}, #3-#4 (#1)}
\newcommand{\yprlN}[4]{, #4, {Phys.\ Rev.\ Lett.\ }{\bf #2}, #3 (#1)}

\newcommand{\sapj}[3]{, #2, {Astrophys.\ J.}, submitted, arXiv:#3 (#1)}
\newcommand{\yapj}[5]{, #5, {Astrophys.\ J.\ }{\bf #2}, #3-#4 (#1)}

\newcommand{\yapjN}[4]{, #4, {Astrophys.\ J.\ }{\bf #2}, #3 (#1)}
\newcommand{\yapjlN}[4]{, #4, {Astrophys.\ J.\ Lett.\ }{\bf #2}, #3 (#1)}

\newcommand{\yapjl}[5]{, #5, {Astrophys.\ J.\ Lett.\ }{\bf #2}, #3-#4 (#1)}

\newcommand{\yaraa}[5]{, #5, {Ann.\ Rev.\ Astron.\ Astrophys.\ }{\bf #2}, #3-#4 (#1)}

\newcommand{\yanarN}[4]{, #4, {Astron.\ Astrophys.\ Rev.\ }{\bf #2}, #3 (#1)}

\newcommand{\ypf}[5]{, #5, {Phys.\ Fluids }{\bf #2}, #3-#4 (#1)}

\newcommand{\ygafd}[5]{, #5, {Geophys.\ Astrophys.\ Fluid Dynam. }{\bf #2}, #3-#4 (#1)}

\newcommand{\yjour}[6]{, #6, {#2} {\bf #3}, #4-#5 (#1)}
\newcommand{\yjourN}[5]{, #5, {#2} {\bf #3}, #4 (#1)}

\newcommand{\yproc}[7]{, #4, In {#5} (ed.\ #6), pp.\ #2-#3.\ #7 (#1)}
\newcommand{\ybook}[3]{ {#2}.\ #3 (#1)}

\begin{document}

\title*{Simulations of galactic dynamos}
\author{Axel Brandenburg}
\institute{Axel Brandenburg \at
Nordita, KTH Royal Institute of Technology and Stockholm University,
Roslagstullsbacken 23, SE-10691 Stockholm, Sweden,
\email{brandenb@nordita.org}, $ $Revision: 1.51 $ $}
\maketitle

\abstract{
We review our current understanding of galactic dynamo theory,
paying particular attention to numerical simulations both of the
mean-field equations and the original three-dimensional equations
relevant to describing the magnetic field evolution for a turbulent flow.
We emphasize the theoretical difficulties in explaining non-axisymmetric
magnetic fields in galaxies and discuss the observational
basis for such results in terms of rotation measure analysis.
Next, we discuss nonlinear theory, the role of magnetic helicity
conservation and magnetic helicity fluxes.
This leads to the possibility that galactic magnetic fields may be bi-helical,
with opposite signs of helicity and large and small length scales.
We discuss their observational signatures and close by discussing
the possibilities of explaining the origin of primordial magnetic fields.
}

\section{Introduction}
\label{sec:1}

We know that many galaxies harbor magnetic fields.
They often have a large-scale spiral design.
Understanding the nature of those fields was facilitated by an
analogous problem in solar dynamo theory, where large-scale
magnetic fields on the scale of the entire Sun were explained
in terms of mean-field dynamo theory.
Competing explanations in terms of primordial magnetic fields
have been developed in both cases, but in solar dynamo theory
there is the additional issue of an (approximately) cyclic variation,
which is not easily explained in terms of primordial fields.

Historically, primordial magnetic fields were considered a serious contender
in the explanation of the observed magnetic fields in our and other
spiral galaxies; see the review of \cite{SFW86}.
The idea is simply that the differential rotation of the gas in galaxies
winds up an ambient magnetic field to form a spiraling magnetic field pattern.
There are two problems with this interpretation.
Firstly, if there was no turbulent diffusion, the magnetic field would be
wound up too many times to explain the observed field, whose magnetic spiral
is not as tightly wound as one would have otherwise expected.
The tight winding can be alleviated by turbulent diffusion, which is clearly
a natural process that is expected to occur in any turbulent environment.
In galaxies, an important source of turbulence is supernova explosions
\citep{KBSTN98,Gent13a,Gent13b} that are believed to sustain the canonical
values of a root-mean-square turbulent velocity of $\urms=10\kms$ at
density $\rho=2\times10^{-24}\g\cm^{-3}$.
The required vertically integrated energy input would be of the order
of $0.5\rho\urms^3\approx10^{-24}\g\cm^{-3}\,(10^6\cm\s^{-1})^3
=10^{-6}\erg\cm^{-2}\s^{-1}$.
which is easily balanced by about 20 supernovae with $10^{51}\erg$
per million years per $\kpc^2$ for the solar neighborhood, which yields
about $0.7\times10^{-4}\erg\cm^{-2}\s^{-1}$, i.e., nearly two orders
of magnitude more than needed \citep{BN11}.

Turbulence with these values of $\urms$ and a correlation length
$\ell=70\pc$ \citep{Shu05}, corresponding to a correlation wavenumber
$\kf=2\pi/\ell$ is expected to produce turbulent diffusion
with a magnetic diffusion coefficient
$\etat=\urms/3\kf\approx10^{25}\cm^2\s^{-1}\approx0.04\,\kpc\km\s^{-1}$,
which would lead to turbulent decay of a magnetic field with a vertical
wavenumber of $k=2\pi/0.3\kpc\approx20\kpc^{-2}$ on a decay time of about
$(\etat k^2)^{-1}\approx60\Myr$.
Thus, to sustain such a field, a dynamo process is required.

Magnetic fields affect the velocity through the Lorentz force.
However, if one only wants to understand the origin of the
magnetic field, we would be interested in early times when the
mean magnetic field is still weak.
In that case we can consider the case when the velocity field is still
unaffected by the magnetic field and it can thus be considered given.
This leads to a kinematic problem that is linear.

Several dynamo processes are known.
Of particular relevance are large-scale dynamos that produce
magnetic fields on scales large compared with the size of the
turbulent eddies.
These dynamos are frequently being modeled using mean-field
dynamo theory, which means that one solves the averaged
induction equation.
In such a formulation, the mean electromotive force resulting
from correlations of small-scale velocity and magnetic field
fluctuations are being parameterized as functions of the
mean magnetic field itself, which leads to a closed system
of equations.
The resulting mean-field equations can have exponentially
growing or decaying solutions.
Of particular interest is here the question regarding the
symmetry properties of the resulting magnetic field.
This aspect will be discussed in \Sec{sec:LinTheo}.

Next, the magnetic field will eventually be subject to
nonlinear effects and saturate.
The most primitive form of nonlinearity is $\alpha$ quenching,
which limits the $\alpha$ effect such that the energy density
of the local mean magnetic field strength is of the order of
the kinetic energy of the turbulence.
There is the possibility of so-called catastrophic quenching,
which has sometimes been argued to suppress not only turbulent
diffusion \citep{CV91}, but also the dynamo effect \citep{VC92}.
Those aspects will be discussed in \Sec{sec:NonlinTheo}.
It is now understood that catastrophic quenching is a consequence
of magnetic helicity conservation and the fact that the magnetic
field takes the form of a bi-helical field with magnetic helicity
at different scales and signs.
Such a field might have observational signatures that could
be observable, as will be discussed in \Sec{sec:ObsAspects}.

An entirely different alternative is that a primordial
magnetic field might still exist.
It would be of interest to find out what effects it would have.
This affects the discussion of the initial turbulent
magnetic field, which might occur in conjunction with magnetic helicity,
which requires some knowledge about turbulent cascades that we shall
also discuss in connection with dynamos, so we postpone the discussion
of primordial magnetic fields until \Sec{sec:Primordial}, and
begin with kinematic mean-field theory.

\section{Aspects of kinematic mean-field theory}
\label{sec:LinTheo}

The purpose of this section is to review some of the important results
in applying mean-field dynamo theory to galaxies.
We focus here on linear models and postpone the discussion of
essentially nonlinear effects to \Sec{sec:NonlinTheo}.

\subsection{Dominance of quadrupolar modes}

\begin{figure}[t!]\begin{center}
\includegraphics[width=\textwidth]{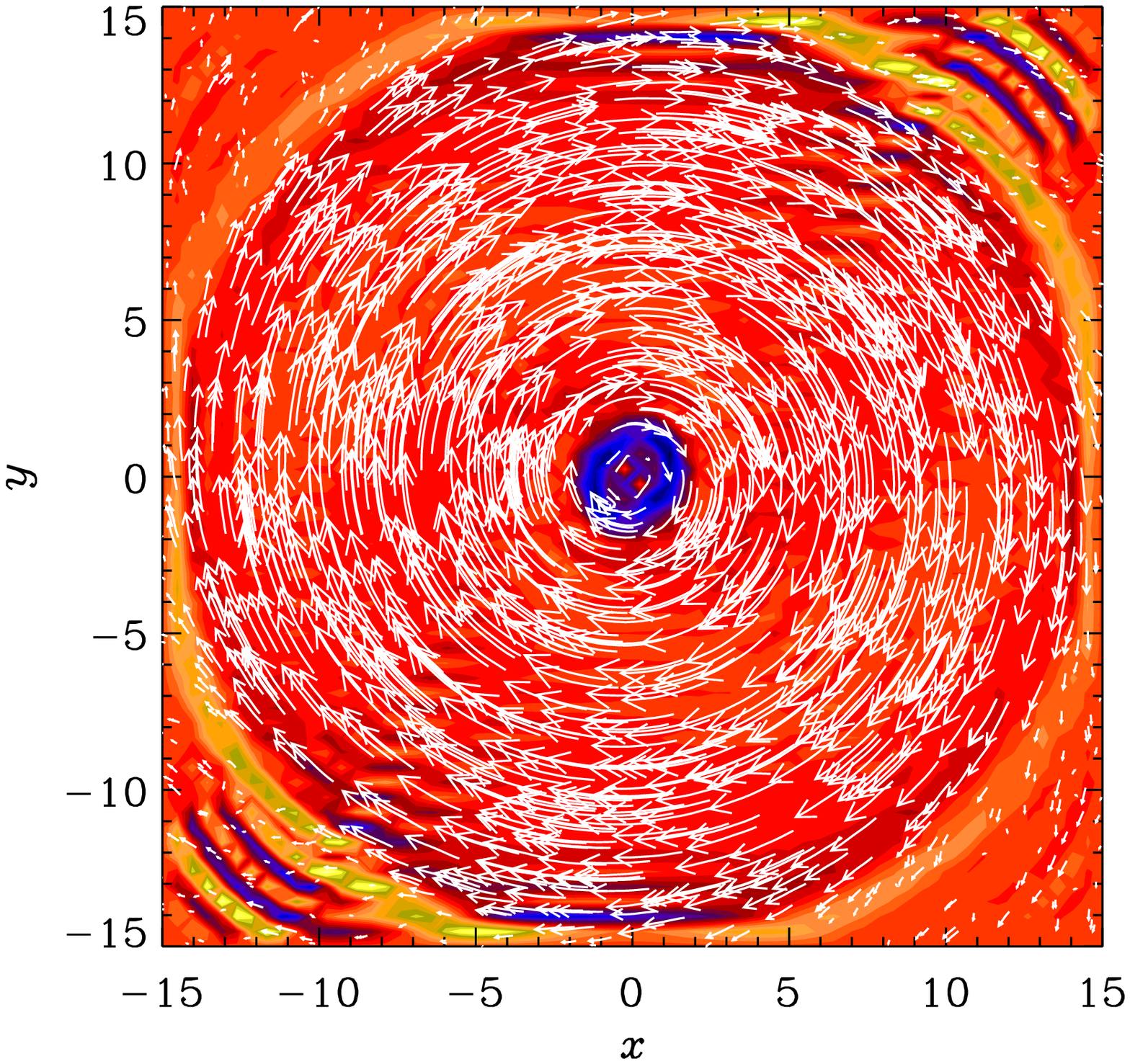}
\includegraphics[width=.87\textwidth]{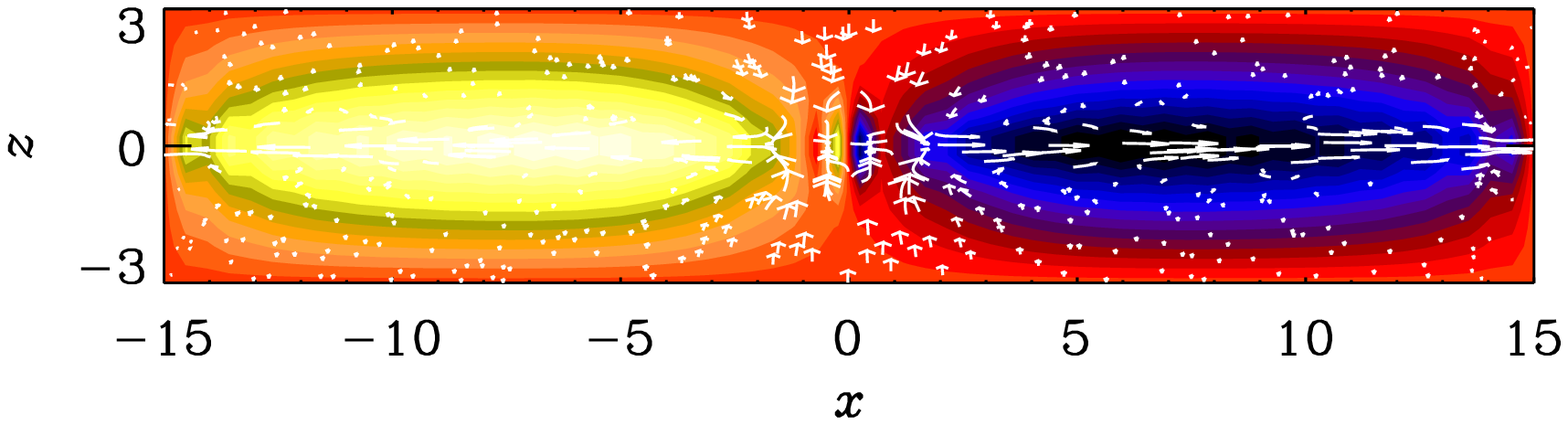}
\end{center}\caption[]{
Magnetic field in the midplane of a simplified model of a
galaxy with $\alpha$ effect and Brandt rotation curve.
}\label{pvar_xy_64x64_15kc_c}\end{figure}

Mean-field dynamo theory for galaxies \citep{Par71,VR71} was developed
soon after the corresponding theory for solar, stellar and planetary
dynamos was first proposed \citep{Par55,SKR66,SK69,SK69b}.
The main difference to stellar dynamos is the flat geometry.
An important consequence of this is the finding that the lowest eigenmode
is of quadrupolar type, which means that the toroidal magnetic field
has the same direction on both sides of the midplane.
An example of this is shown in \Fig{pvar_xy_64x64_15kc_c}, where we show
vectors of the magnetic field in the $xy$ plane of the galactic disc
together with a $xz$ section approximately through the disc axis.
We note that this model has been calculated in Cartesian geometry,
which leads to minor artifacts as can be seen in two corners.

The models in galactic geometry made use of the fact that in flat
geometries, derivatives in the vertical ($z$) direction are much more
important than in the radial or azimuthal directions.
One therefore deals essentially with one-dimensional models of the form
\citep{RSS88},
\EQ
\dot{\meanB}_R = -(\alpha\meanB_{\phi})' + \eta_{\rm T} \meanB_R'', \quad
\dot{\meanB}_{\phi} = S\meanB_R + \eta_{\rm T} \meanB_{\phi}''.
\label{discdyn}
\EN
Here, primes and dots denote $z$ and $t$ derivatives, respectively,
$\alpha=\alpha_0 f_\alpha(z)$ is a profile for $\alpha$ (asymmetric
with respect to $z=0$) with typical value $\alpha_0$,
$S = R\dd\Omega/\dd R$ is the radial shear in the disc,
and $(\meanB_R,\meanB_{\phi},\meanB_z)$ are the components
of the mean field $\meanBB$ in cylindrical coordinates.
On $z=\pm H$ one assumes vacuum boundary conditions which, in this
one-dimensional problem, reduce to $\meanB_R=\meanB_\phi=0$.
One can also impose boundary conditions on the mid-plane, $z=0$,
by selecting either symmetric (quadrupolar) fields, $B_R=\meanB_\phi'=0$,
or antisymmetric (dipolar) fields, $B_R'=\meanB_\phi=0$.
We define two dimensionless control parameters,
\EQ
C_\Omega = S H^2/\eta_{\rm T}, \quad
C_\alpha = \alpha_0 H/\eta_{\rm T},
\EN
which measure the strengths of shear and $\alpha$ effects, respectively.

\begin{figure}[t!]\begin{center}
\includegraphics[width=\textwidth]{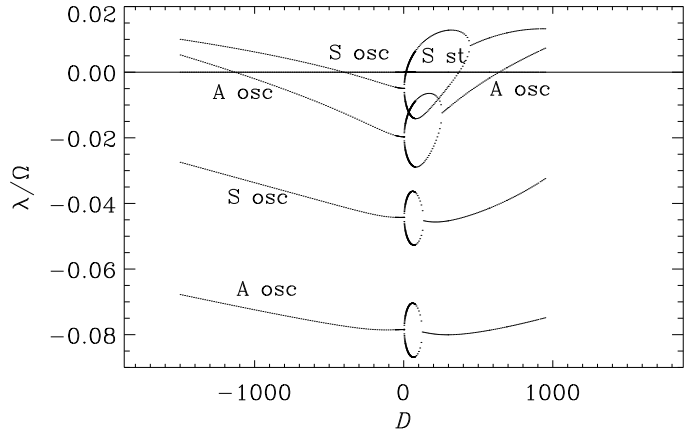}
\end{center}\caption[]{
Eigenvalues of the dynamo equations with radial shear in slab geometry.
The dynamo number is defined positive when shear is negative and
$\alpha$ positive.
Note that for $\alpha>0$, the most easily excited solution is
non-oscillatory (`steady') and has even parity (referred to as
`S~st') while for $\alpha<0$ it is oscillatory (`S~osc').
Adapted from \cite{B98}.
}\label{weigval_vertfield}\end{figure}

In the limit of strong differential rotation, $C_\alpha/C_\Omega\ll1$,
the solutions are characterized by just one parameter, the dynamo number
$D=C_\alpha C_\Omega$.
\FFig{weigval_vertfield} shows the growth rate of different
modes, obtained by solving \Eq{discdyn}
for both signs of the dynamo number \citep{B98}.
To find all the modes, even the unstable ones, one can
solve \Eq{discdyn} numerically as an eigenvalue problem, where the
complex growth rate $\lambda$ is the eigenvalue with the largest real part.
Note that the most easily excited mode is quadrupolar and non-oscillatory.
We denote it by `S~st', where `S' refers to symmetry about the
midplane and `st' refers to steady or as opposed to oscillatory.
Of course, only in the marginally excited case those modes are steady.

The basic dominance of quadrupolar magnetic fields is also reproduced
by more recent global simulations such as those of \cite{GFD09}.
However, the situation might be different in so-called cosmic-ray driven
dynamos (see below), where the magnetic field could be preferentially
dipolar with a reversal of the toroidal play about the midplane
\citep{HOKL09}.
The possibility of a significant dipolar component has also been
found for the magnetic field of our Galaxy \citep{JF12}.
On the other hand, in the inner parts of galaxy, the geometrical
properties of the bulge may also give rise to a locally dipolar
field in the center \citep{DB90}.

\subsection{Non-axisymmetric magnetic fields}
\label{NonAxi}

\begin{figure}[t!]\begin{center}
\includegraphics[width=.9\textwidth]{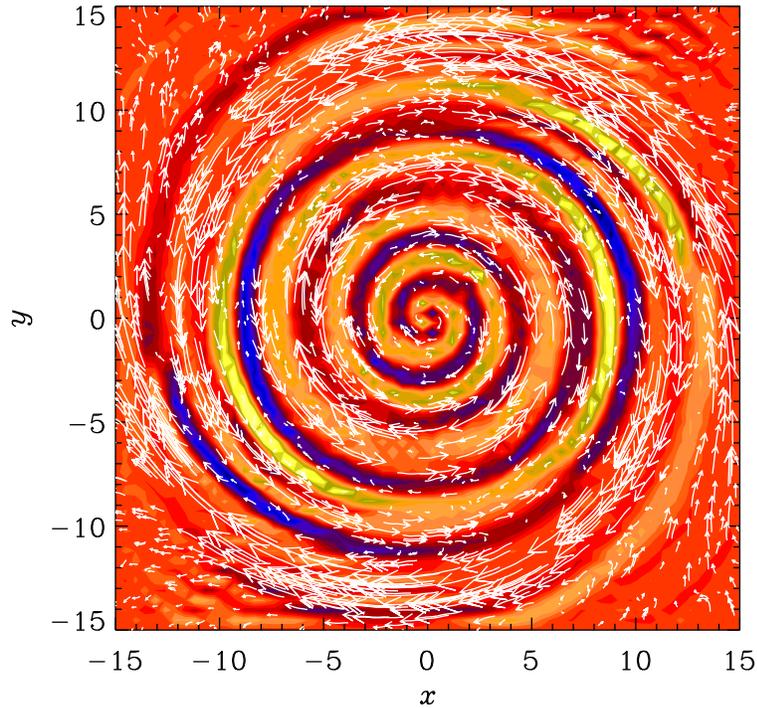}
\end{center}\caption[]{
Magnetic field in the midplane of a simplified model of a
galaxy with Brandt rotation curve and an initially horizontal
magnetic field in the $x$ direction that is then being wound up.
}\label{pvar_xy_64x64_15kc_b}\end{figure}

An important realization due to \cite{Rad86} is that non-axisymmetric
solutions are never favored by differential rotation, because it winds up
such fields, so anti-parallel field lines are being brought close together
and then decay rapidly, as can be seen from \Fig{pvar_xy_64x64_15kc_b}.
This was already found in earlier numerical eigenvalue calculations
\citep{Rae80,Rae86}, suggesting that corresponding asymptotic calculations
that make the so-called $\alpha\Omega$ approximation \citep{RSS85},
in which the $\alpha$ effect is neglected compared with the shear term,
could be problematic.

\begin{figure}[t!]\begin{center}
\includegraphics[width=\textwidth]{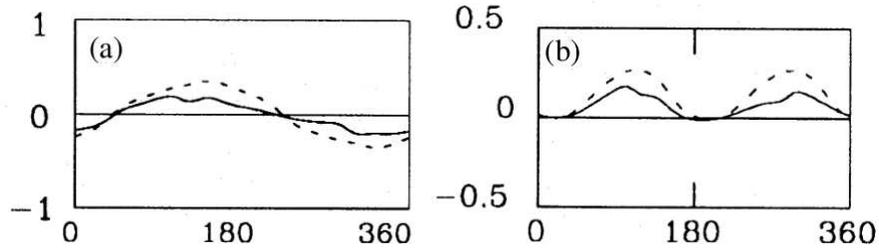}
\end{center}\caption[]{
Dependence of $\RM$ on the azimuthal angle for
(a) magnetic field of type S0, and (b) magnetic field of type S1.
Solid and dashed lines refer to two different procedures of measuring $\RM$.
Adapted from \cite{DB90}.
}\label{RM_S0_S1}\end{figure}

At the time it was thought that many external galaxies would harbor
non-axisymmetric magnetic fields \citep{SFW86}, but this view
has now changed with the more careful measurements of the toroidal
magnetic fields along an azimuthal ring around various external galaxies.
This can be seen from plots of the rotation measure at different
positions along such an azimuthal ring.
The rotation measure is defined as
\EQ
\RM=\dd\chi/\dd\lambda^2,
\EN
where $\lambda$ is the wavelength of the radio emission and
$\chi$ is the angle of the polarization vector determined
from the Stokes parameters $Q$ and $U$ as
\EQ
\chi=\half\Atan(U,Q),
\EN
where $\Atan$ returns all angles between $-\pi$ and $\pi$ whose tangent
yields $U/Q$.
\FFig{RM_S0_S1} shows theoretical $\RM$ dependencies on azimuth
around projected rings around dynamo models simulating galactic magnetic
fields of types `S0' (symmetric about midplane with $m=0$) and `S1'
(also symmetric about midplane, but with $m=1$, i.e., non-axisymmetric).
The result is quite clear.
When the magnetic field is axisymmetric, one expects the toroidal
magnetic field to give a line-of-sight component $B_\|$
and point toward the observer at one azimuthal position
on the projected azimuthal ring and away from the observer on the
opposite position along the ring.
This should lead to a sinusoidal modulation of $\RM$ with one positive
extremum and one negative one.
When the field is bisymmetric, i.e., non-axisymmetric with $m=1$),
one expects two positive extrema and two negative ones.
This is indeed borne out by the simulations.
It is this type of evidence that led to the conclusion that the
magnetic field of M81 is non-axisymmetric \citep{KBH89}.

Even today, M81 is still the one and only example of a galaxy
displaying a distinctly non-axisymmetric magnetic field with
azimuthal order $m=1$ \citep{BBMSS96}; see also the chapter
by Beck on observations of galactic magnetic fields.
Such fields are hard to explain theoretically, because, according
to most of the dynamo models presented so far, non-axisymmetric modes
are always harder to excite than axisymmetric ones; see also
\cite{BTR89} for a survey of such solutions.
The currently perhaps best explanation for non-axisymmetric magnetic fields
in galaxies is that they are a left-overs from the initial conditions
and are being wound up by the differential rotation.
This can be a viable explanation only because for galaxies the
turbulent decay time might be slow enough, especially in their outer parts,
if those fields are helical and non-kinematic \citep{BS13,BBS13}.
\cite{MBDT93} presented a model that incorporated a realistic
representation of the so-called peculiar motions of M81 that
were proposed to be the result of a recent close encounter with a
companion galaxy.
These peculiar motions are flows relative to the systematic differential
rotation and have been obtained from an earlier stellar dynamics
simulation by \cite{TD93}.
A very different alternative is that the $m=1$ magnetic fields in the
outskirts of M81 is driven by the magneto-rotational instability, as
has recently been proposed by \cite{GEZ13}.

A more typical class of non-axisymmetric fields are those with
$m=2$ and $m=0$ contributions.
Those would no longer be called bisymmetric and fall outside the
old classification into axisymmetric and bisymmetric spirals.
Examples of non-axisymmetric but non-bisymmetric spirals are
NGC 6946 and IC 342 \citep[e.g.][]{Beck07,BW13}.
A natural way of explaining such fields is via a non-axisymmetric
dynamo parameters such as the $\alpha$-effect \citep{MS91,MBT91,SM93}.
This has been confirmed through more realistic modeling both with
\citep{CSS13,CSQ14} and without \citep{MBSSKA13} memory effect.
For a more popular account on recent modeling efforts, see also
the review by \cite{Moss12}.

\subsection{The $\alpha$ effect and turbulent diffusivity in galaxies}

The forcing of turbulence through the pressure force associated with
the thermal expansion of blast waves is essentially irrotational.
However, vorticity is essential for what is known as small-scale dynamo
action, and it is also a defining element of kinetic helicity and hence
also the $\alpha$ effect, which is 
an important parameter in mean-field simulations of galactic dynamos.
In galaxies, the baroclinic term is an important agent for making the
resulting flow vortical \citep{KBT98}; see also \cite{DSB11},
who also compared with the effects of rotation and shear.
Thus, we need to know how efficiently vorticity can be generated
in turbulence.
This question becomes particularly striking in the case of isothermal
turbulence, because then, and in the absence of rotation, shear, or
magnetic fields, there is no baroclinic term that could generate vorticity.
In that case, vorticity can only be generated via viscosity through
a ``visco-clinic'' term of the form $\nab\ln\rho\times\nab\dive\uu$,
although it is not obvious that this term is unaffected by the numerical
form of the diffusion operator.

\begin{figure}[t!]\begin{center}
\includegraphics[width=\textwidth]{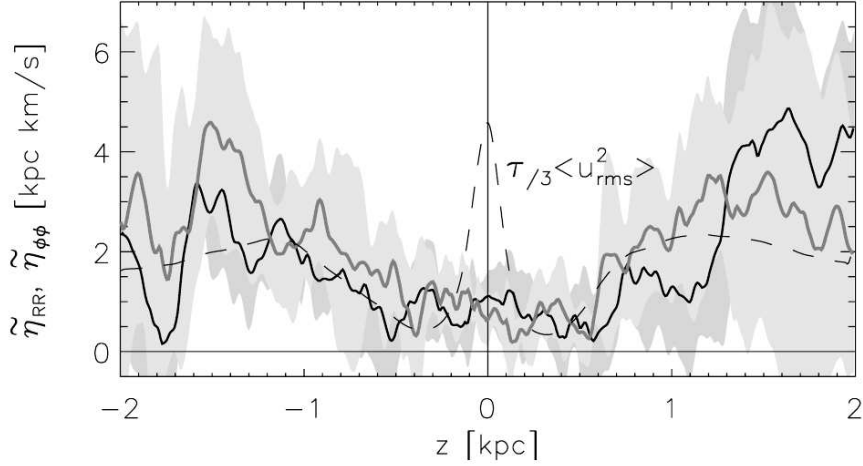}
\end{center}\caption[]{
Vertical dependence of $\etat$ obtained by the test-field method
using a simulation of supernova-driven turbulence.
Adapted from \cite{GZER08}.
}\label{gressel08}\end{figure}

Most of the papers assume that it is a result of cyclonic turbulent
motions driven by supernova explosions.
There have also been attempts to calculate $\alpha$ and $\etat$ by considering
individual explosions \citep{Fer92} and also so-called superbubbles
resulting from several explosions that could have triggered each other
\citep{Fer93}.

Nowadays, a reliable method for calculating $\alpha$ and $\etat$ from
numerical simulations is the test-field method, which will be
briefly discussed below.
The $\alpha$ effect and turbulent diffusivity $\etat$ characterize
the resulting electromotive force $\meanEMF$ from small-scale (unresolved)
motions, i.e., $\meanEMF=\overline{\uu\times\bb}$.
This expression enters in the evolution equation for the mean magnetic field,
\EQ
{\partial\meanBB\over\partial t}=\nab\times\left(
\UU\times\BB+\overline{\uu\times\bb}-\eta\mu_0\meanJJ\right).
\label{dmeanBBdt}
\EN
To determine $\overline{\uu\times\bb}$ as a function of $\meanBB$,
which drives magnetic fluctuations $\bb=\BB-\meanBB$ through tangling
by the turbulent motions $\UU=\meanUU+\uu$, we use the evolution equation
for $\bb$ obtained by subtracting \Eq{dmeanBBdt} from the full
induction equation
\EQ
{\partial\BB\over\partial t}=\nab\times\left(
\UU\times\BB-\eta\mu_0\JJ\right),
\label{dBBdt}
\EN
with the result
\EQ
{\partial\bb\over\partial t}=\nab\times\left(\meanUU\times\bb+\uu\times\meanBB
+\uu\times\bb-\overline{\uu\times\bb}-\etat\mu_0\jj\right).
\label{dbbdt}
\EN
We solve this equation for mean fields $\meanBB$ that do not need
to be a solution of \Eq{dmeanBBdt}; see \cite{Sch05,Sch07}.
We can then compute $\overline{\uu\times\bb}$, related it to the
chosen test fields and their derivatives, $\meanB_i$ and
$\partial\meanB_i/\partial x_j$, and determine all relevant
components of $\alpha_{ij}$ and $\eta_{ijk}$, which requires
a corresponding number of test fields.

There is by now a lot of literature on this topic.
The method has been extended into the quasi-kinematic \citep{BRRS08}
and fully nonlinear \citep{RB10} regimes.
For moderate scale separation, a convolution in space and time
can often not be ignored \citep{BRS08,HB09}, but it is possible
to incorporate such effects in an approximate fashion by solving
an evolution equation for the mean electromotive force \citep{RB12}.
Such an approach restores causality in the sense that the elliptic
nature of the diffusion equation takes the form of a wave equation,
which limits effectively the maximum propagation speed to the
rms velocity of the turbulence \citep{BKM04}.
Such an equation is usually referred to as the telegraph equation.
In galaxies, such an effect can also cause magnetic arm to lag the
corresponding material arm with respect to the rotation \citep{CSS13}.

\cite{GZER08} have applied the test-field method to their turbulent
galactic dynamo simulations \citep{GEZR08} and find values for $\etat$
that are of the order of $1\kpc\km\s^{-1}$ (see \Fig{gressel08}), which
corresponds to $3\times10^{26}\cm^2\s^{-1}$, which is 30 times
larger than our naive estimate presented in the introduction.
In their simulations, $\urms\approx40\kms$, so their effective
value of $\kf$ must be $\kf\approx\urms/3\etat\approx13\kpc^{-1}$,
and their effective correlation length thus $2\pi/\kf\approx0.5\kpc$,
instead of our estimate of only $0.07\kpc$.
The reason for this discrepancy in unclear and highlights
the importance of doing numerical simulations.
Their values for $\alpha$ are positive in the upper disc plane and
increase approximately linearly to about $5\km\s^{-1}$ at a height
of $1\kpc$.
This allows us to estimate the fractional helicity as
$\epsf\approx\alpha/\etat\kf\approx0.1$ \citep[cf.][]{BB02}.

An additional driver of the $\alpha$ effect is the possibility of
inflating magnetic flux tubes by cosmic rays \citep{Par92}.
This makes such magnetic flux tubes buoyant and, together with the
effects of rotation and stratification, leads to an $\alpha$ effect.
This has led to successful simulations of galactic dynamos both
in local \citep{HKOL04} and global \citep{HOKL09,Kulpa11} geometries.
Cosmic rays are usually treated in the diffusion approximation with
a diffusion tensor proportional to $B_i B_j$ that forces the diffusion
to be only along magnetic field lines.
However, the effective diffusivity is very large
(in excess of $10^{28}\cm^2\s^{-1}$), making an explicit treatment
costly because of a short diffusive time step constraint.
Again, one can make use of the telegraph equation to limit the diffusion
speed to a speed not much faster than the speed of sound.
Such an approach has been exploited by \cite{SBMS06}, but it has not
yet been applied to more realistic cosmic ray-driven dynamos.

\section{Aspects of nonlinear mean-field theory}
\label{sec:NonlinTheo}

\subsection{Bi-helical magnetic fields from simulations}

\begin{figure}[t!]\begin{center}
\includegraphics[width=\columnwidth]{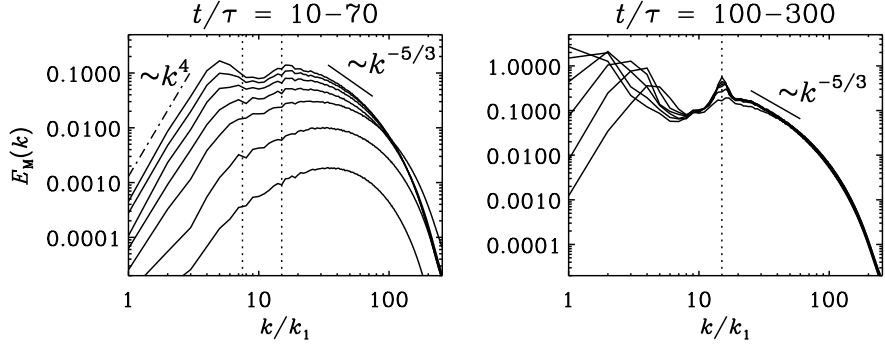}
\end{center}\caption[]{
Magnetic energy spectra $\EM(k)$, at earlier (left) and later (right) times.
The scale separation ratio is $\kf/k_1=15$.
The range of time $t$ is given in  units of the turnover time,
$\tau=1/\urms\kf$.
At small wavenumbers, the $\EM(k)$ spectrum is proportional to $k^4$,
while to the right of $\kf/k_1=15$ there is a short range with
a $k^{-5/3}$ spectrum.
Adapted from \cite{B11}.
}\label{pspec_ck_bete12}\end{figure}

When the computational domain is large enough and turbulence
is driven in a helical fashion at a small length scale,
one sees the clear emergence of what is called a bi-helical
magnetic field.
An example is shown in \Fig{pspec_ck_bete12} where we show
magnetic power spectra of a simulation of \cite{B11}.
During the early evolution of the dynamo (left) we see the
growth of the magnetic field at small wavenumbers,
accompanied by a growth at small amplitude at lower wavenumbers.
The spectrum remains however roughly shape-invariant.

By $t/\tau>100$, where $\tau$ is the turnover time of the turbulence
at the forcing scale, a large-scale field is already present.
As the field saturates, the peak of magnetic energy moves to progressively
smaller wavenumbers.
The reason for this peak can be understood in terms of an
$\alpha^2$ dynamo.

\subsection{Catastrophic quenching}

The idea of catastrophic quenching is almost as old as mean-field
dynamo theory itself.
Here, ``catastrophic'' refers to a (declining) dependence of the turbulent
transport coefficients on the magnetic Reynolds number, $\Rm$, even when
$\Rm$ is very large.
The word `catastrophic' was first used by \cite{BF00} to indicate
the fact that in the astrophysical context, the $\alpha$ effect
would become catastrophically small.
The issue focussed initially on the issue of turbulent diffusion
\citep{Kno78,Lay79,Pid81}.
Numerical simulations later showed that in two dimensions,
with a large-scale magnetic field lying in that plane,
the decay of this large-scale field is indeed slowed down
in an $\Rm$-dependent (i.e., catastrophic) fashion \cite{CV91}.
This then translates into a corresponding $\Rm$-dependent quenching
of the effective turbulent magnetic diffusivity.
Later, \cite{VC92} argued that also the $\alpha$ effect would be
catastrophically quenched, possibly with an even higher power of $\Rm$.

\cite{GD94} later realized that the $\Rm$ dependence is associated
with the presence of certain conservation laws which are different
in two and three dimensions.
In three dimensions, the magnetic helicity, $\bra{\AAA\cdot\BB}$,
is conserved, while in two dimensions, $\bra{A^2}$ is conserved.
Here and elsewhere, angle brackets denote averaging, and $A$ is the
component of $\AAA$ that is perpendicular to the plane in two dimensions.

In three dimensions, the suppression of the large-scale dynamo effect
can be understood by considering the fact that the field generated by
an $\alpha$-effect dynamo is helical and of Beltrami type, e.g.,
$\meanBB=(\sin k_1z,\cos k_1z,0)B_0$, which is parallel to its curl, i.e.,
$\nab\times\meanBB=(\sin k_1z,\cos k_1z,0)kB_0=k_1\meanBB$.
The vector potential can then be written as $\meanAA=k_1^{-1}\meanBB$, so that
the magnetic helicity is $\bra{\meanAA\cdot\meanBB}=k_1^{-1}B_0^2$.
The current helicity is $\bra{\meanJJ\cdot\meanBB}=k_1 B_0^2/\mu_0$.
Note, however, that magnetic helicity of the total field is conserved.
Since the total field is given by the sum of large and small-scale components,
$\BB=\meanBB+\bb$, the generation of magnetic helicity at large scales can be
understood if there is a corresponding production of magnetic helicity at
small scales, but of opposite sign.
Writing for the magnetic and current helicities of the small-scale field
analogously $\bra{\aaaa\cdot\bb}=-\kf^{-1}\bra{\bb^2}$ and
$\bra{\jj\cdot\bb}=-\kf\bra{\bb^2}/\mu_0$.

\begin{figure}[t!]\begin{center}
\includegraphics[width=\columnwidth]{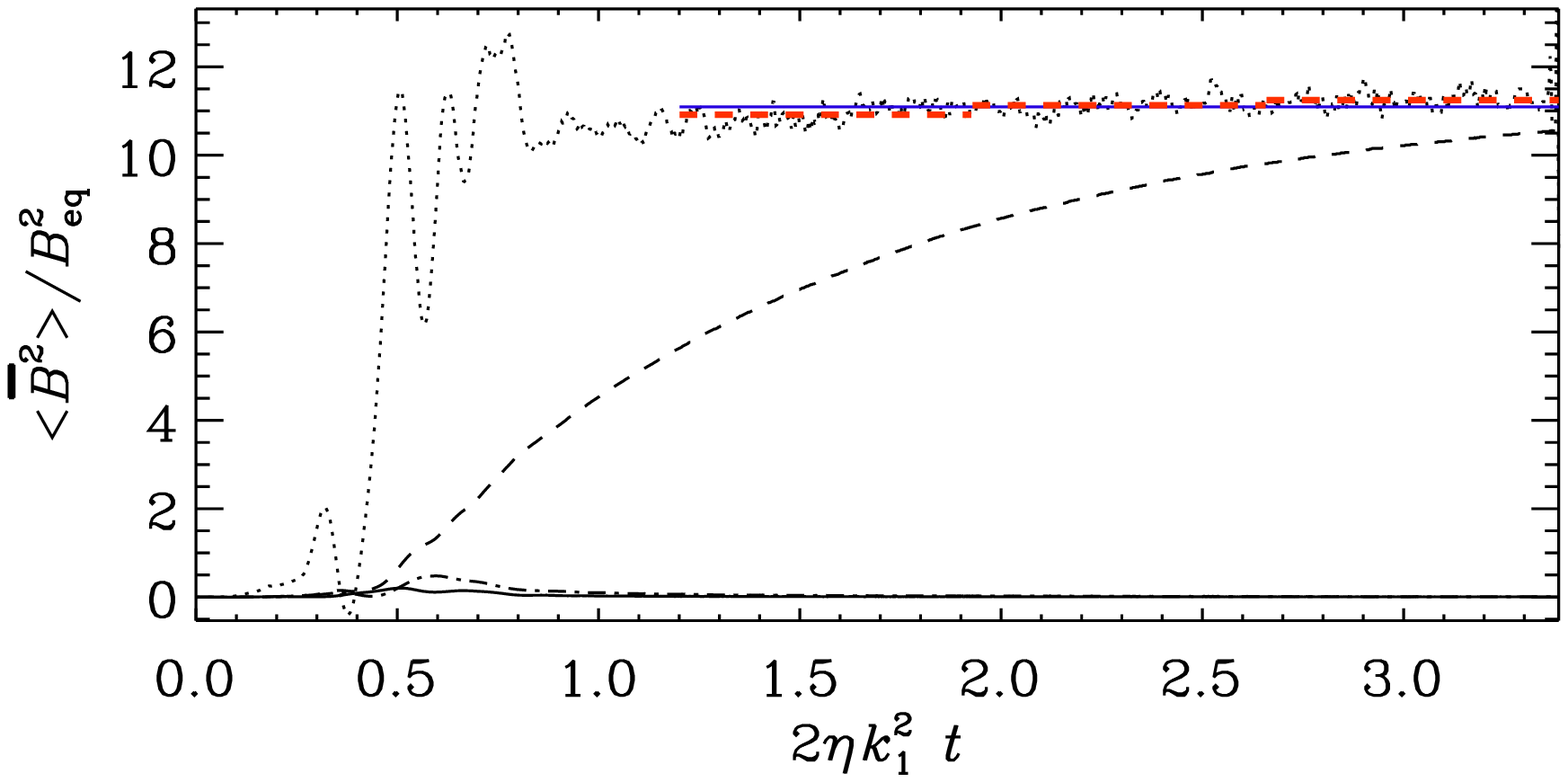}
\end{center}\caption[]{
Example showing the evolution of the normalized $\bra{\meanBB^2}$ (dashed)
and that of $\bra{\meanBB^2}+\dd\bra{\meanBB^2}/\dd(2\eta k^2t)$ (dotted),
compared with its average in the interval $1.2\leq2\eta k_1^2t\leq3.5$
(horizontal blue solid line), as well as averages over 3 subintervals
(horizontal red dashed lines).
Here, $\meanBB$ is evaluated as an $xz$ average, $\bra{\BB}_{xz}$.
For comparison we also show the other two averages, $\bra{\BB}_{xy}$
(solid) and $\bra{\BB}_{yz}$ (dash-dotted), but their values are very small.
Adapted from \cite{CB13}.
}\label{psat}\end{figure}

The relative importance of large-scale and small-scale contributions
to magnetic helicity and magnetic energy is determined by the magnetic
helicity equation,
\begin{equation}
{\dd\over\dd t}\bra{\AAA\cdot\BB}=-2\eta\mu_0\bra{\JJ\cdot\BB}.
\label{dAB}
\end{equation}
Inserting $\bra{\AAA\cdot\BB}=\bra{\meanAA\cdot\meanBB}+\bra{\aaaa\cdot\bb}$
and $\bra{\JJ\cdot\BB}=\bra{\meanJJ\cdot\meanBB}+\bra{\jj\cdot\bb}$,
and applying it to the time after which the small-scale field has already
reached saturation, i.e., $\bra{\bb^2}=\const$, we have
\begin{equation}
k_1^{-1}{\dd\over\dd t}\bra{\meanBB^2}=-2\eta k_1\bra{\meanBB^2}
+2\eta\kf\bra{\bb^2}.
\label{dBmean}
\end{equation}
One sees immediately that the steady state solution is
$\bra{\meanBB^2}/\bra{\bb^2}=\kf/k_1>1$, i.e., the large-scale
field exceeds the small-scale field by a factor that is equal to
the scale separation ratio.
Moreover, this steady state is only reached on a resistive time scale.
Since $\bra{\bb^2}$ is assumed constant in time, we can integrate \Eq{dBmean}
to give
\begin{equation}
\bra{\meanBB^2}=\bra{\bb^2}{\kf\over k_1}
\left[1-e^{-2\eta k_1^2(t-t_{\rm sat})}\right],
\label{Bmean}
\end{equation}
which shows that the relevant resistive time scale is $(2\eta k_1^2)^{-1}$.
This saturation behavior agrees well with results from simulations;
see \Fig{psat}.

\subsection{Mean-field description}
\label{MeanField}

The simplistic explanation given above can be reproduced in
mean-field dynamo theory when magnetic helicity conservation
is introduced as an extra constraint.
Physically, such a constraint is well motivated and goes back early
work of \cite{PFL}, who found that the relevant $\alpha$ in the
mean-field dynamo is given by the sum of kinetic and magnetic contributions,
\begin{equation}
\alpha=\alphaK+\alphaM,
\label{alphaKM}
\end{equation}
where $\alphaK=-(\tau/3)\bra{\oo\cdot\uu}$ is the formula for
the kinematic value in the high conductivity limit \citep{Mof78,KR80}, and
$\alphaM=(\tau/3\meanrho)\bra{\jj\cdot\bb}$ is the magnetic contribution.
Again, under isotropic conditions, $\bra{\jj\cdot\bb}$ is proportional to
$\bra{\aaaa\cdot\bb}$, with the coefficient of proportionality being $\kf^2$.
This is because the spectra of magnetic and current helicity,
$H(k)$ and $C(k)$, which are normalized such that
$\int H(k)\,\dd k=\bra{\AAA\cdot\BB}$ and $\int C(k)\,\dd k=\bra{\JJ\cdot\BB}$,
are proportional to each other with $C(k)=k^2 H(k)$, so the proportionality
between small-scale current and magnetic helicities is obtained by applying
$C(k)=k^2 H(k)$ to $k=\kf$.
Even in an inhomogeneous system, this approximation is qualitatively
valid, except that the coefficient of proportionality is found to be
somewhat larger \citep{MCCTB10,HB10,DSGB13}.

\begin{figure}[t!]\begin{center}
\includegraphics[width=\columnwidth]{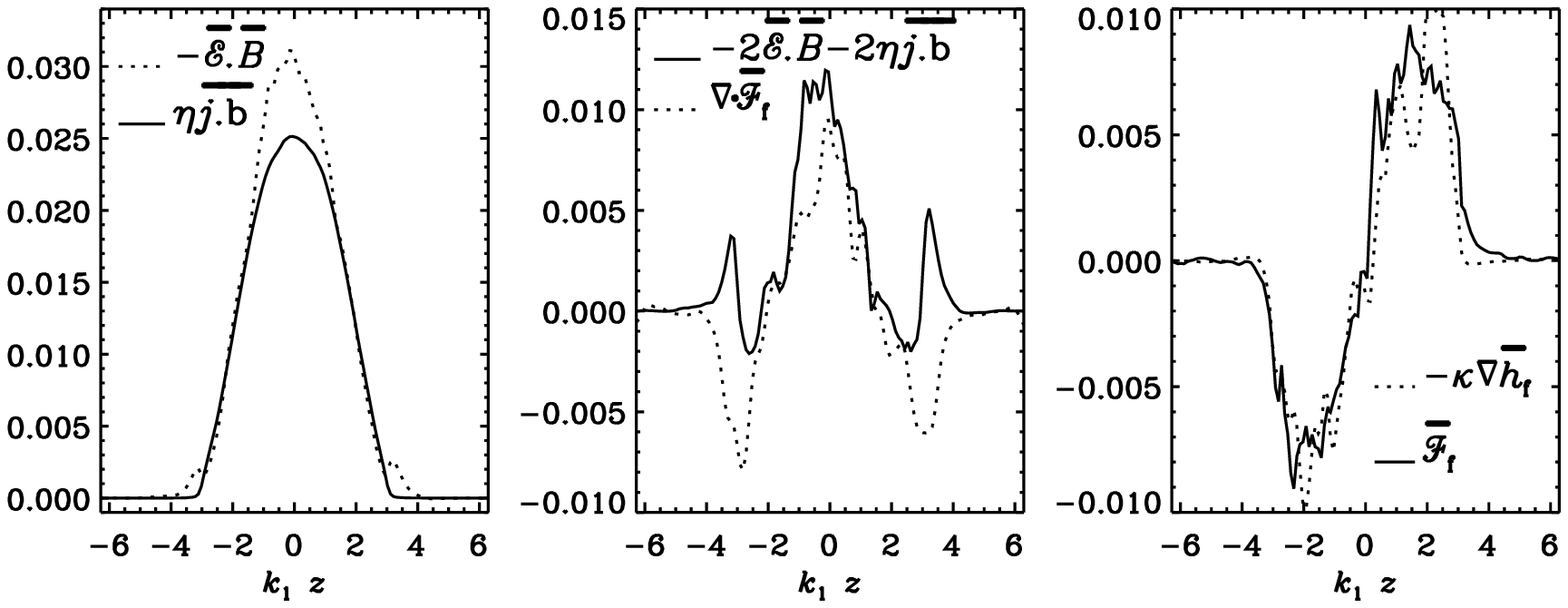}
\end{center}\caption[]{
Time-averaged terms on the right-hand side of \eq{dab},
$\bra{\meanEMF\cdot\meanBB}_T$ and $\eta\bra{\jj\cdot\bb}_T$ (left panel),
the difference between these terms compared with the magnetic helicity flux
divergence of small-scale fields $\bra{\nab\cdot\meanFFFF_{\rm f}^{\rm W}}_T$
(middle panel), and the flux itself compared with the Fickian diffusion
ansatz (right-hand panel).
Adapted from \cite{HB10}.
}\label{pflux_profile}\end{figure}

The question is now how to obtain $\bra{\aaaa\cdot\bb}$.
One approach is to evolve $\meanAA$ (instead of $\meanBB$) in a mean-field
model and compute at each time step \citep{HB12}
$\bra{\aaaa\cdot\bb}=\bra{\AAA\cdot\BB}-\bra{\meanAA\cdot\meanBB}$,
where $\bra{\AAA\cdot\BB}$ obeys \Eq{dAB} for the total magnetic helicity.
In practice, one makes an important generalization in that volume
averaging is relaxed to mean just averaging over one or at most two
coordinate directions.
(To obey Reynolds rules, these coordinate directions should be periodic.)
Thus, \Eq{dAB} then becomes
\begin{equation}
{\partial\over\partial t}\overline{\AAA\cdot\BB}=
-2\eta\mu_0\overline{\JJ\cdot\BB}-\nab\cdot\meanFF,
\label{dABm}
\end{equation}
where $\meanFF$ is the magnetic helicity flux from both large-scale
and small-scale fields.
\cite{HB12} pointed out that this approach can be superior to the
traditional approach by \cite{KR82}, in which one solves instead
the evolution equation for $\bra{\aaaa\cdot\bb}$,
\begin{equation}
{\partial\over\partial t}\overline{\aaaa\cdot\bb}=-2\meanEMF\cdot\meanBB
-2\eta\mu_0\overline{\jj\cdot\bb}-\nab\cdot\meanFFf,
\label{dab}
\end{equation}
where $\meanFFf$ is the magnetic helicity flux only from the small-scale
magnetic field.
The two approaches are equivalent, except that there is an ambiguity as
to what should be included in $\meanFFf$.
In particular, when deriving the evolution equation for
$\bra{\meanAA\cdot\meanBB}$ in the Weyl gauge, i.e., using just
$\partial\meanAA/\partial t=\meanEMF-\eta\mu_0\meanJJ$, we obtain
\begin{equation}
{\partial\over\partial t}(\meanAA\cdot\meanBB)=
2\meanEMF\cdot\meanBB-2\eta\mu_0\meanJJ\cdot\meanBB
-\nab\cdot(-\meanEMF\times\meanAA),
\label{dAmBm}
\end{equation}
i.e., there is an extra flux term $-\meanEMF\times\meanAA$.
Thus, as argued by \cite{HB12}, if $\meanFF=\meanFFm+\meanFFf=\nullvector$,
this implies that $\meanFFf=\meanEMF\times\meanAA$ in \Eq{dAmBm}.

\subsubsection{Diffusive magnetic helicity fluxes and gauge issues}

Another important contribution to the magnetic helicity flux is a
turbulent-diffusive flux down the gradient of magnetic helicity density.
In \Fig{pflux_profile} we show the profiles of
$\bra{\meanEMF\cdot\meanBB}$ and $\eta\bra{\meanJJ\cdot\meanBB}$
from a simulation of \cite{HB10}, compare the residual
$2\bra{\meanEMF\cdot\meanBB}-2\eta\bra{\meanJJ\cdot\meanBB}$
with the divergence of the magnetic helicity flux, and finally
compare the flux $\meanFFFF_{\rm f}=\overline{\ee\times\aaaa}$
with that obtained from the diffusion approximation,
$-\kappa_{\rm f}\nab\meanh_{\rm f}$.
These results demonstrate that there is indeed a measurable difference
between $\bra{\meanEMF\cdot\meanBB}$ and $\eta\bra{\meanJJ\cdot\meanBB}$,
which can be explained by a magnetic helicity flux divergence, and
that this magnetic helicity flux can be understood as a turbulent-diffusive
one, i.e., down the gradient of the local magnetic helicity density.

\begin{figure}[t!]\begin{center}
\includegraphics[width=.8\textwidth]{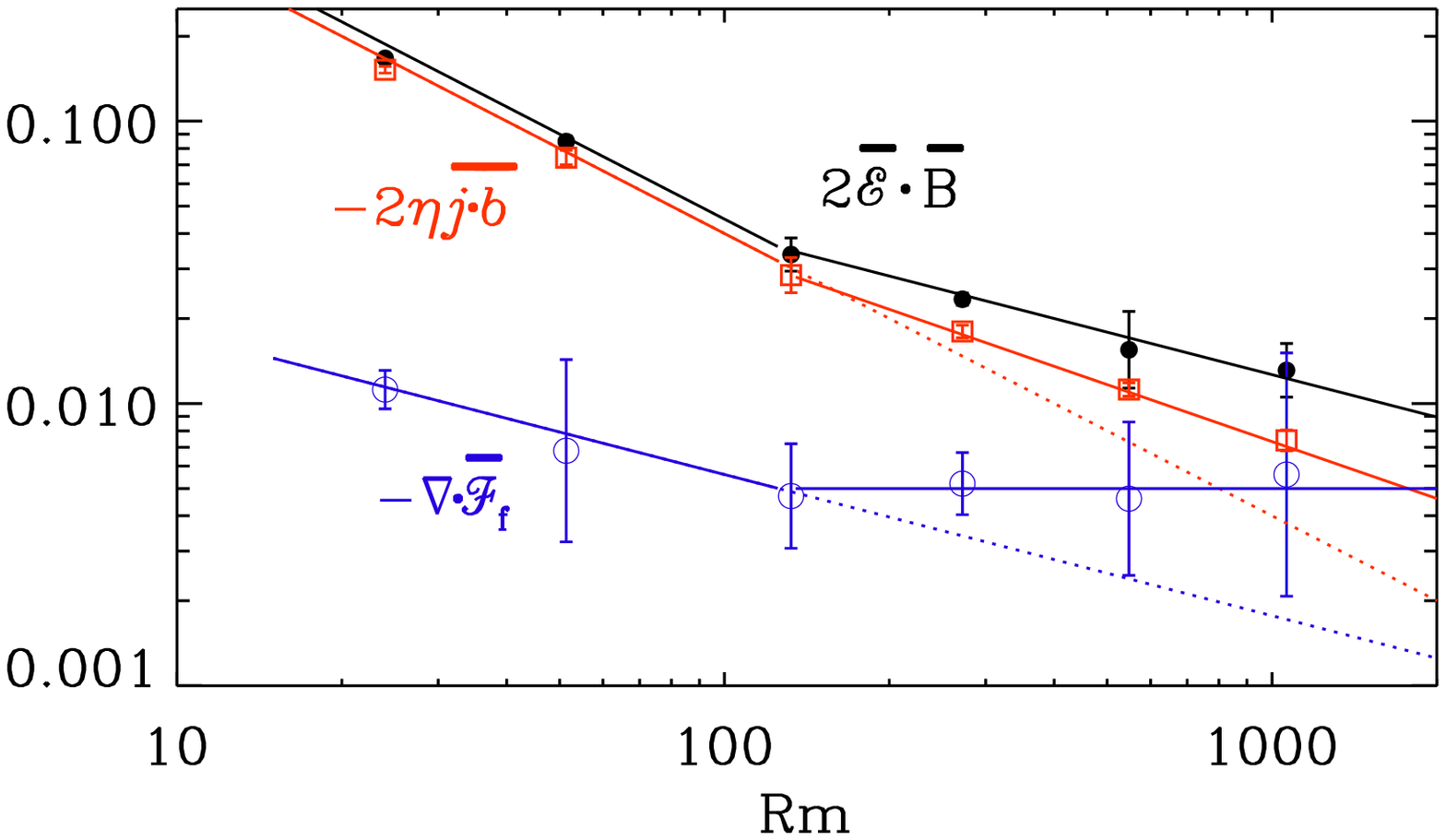}
\end{center}\caption[]{
Scaling properties of the vertical slopes of $2\meanEMF\cdot\meanBB$,
$-2\eta\mu_0\,\overline{\jj\cdot\bb}$, and $-\nab\cdot\meanFFf$ for
models with a wind.
(Given that the three quantities vary approximately linearly with $z$,
the three labels indicate their non-dimensional values at $k_1z=1$.)
The second panel shows that a stronger wind decreases the value of $\Rm$
for which the contribution of the advective term becomes comparable
to that of the resistive term.
Adapted from \cite{DSGB13}.
}\label{presults2}\end{figure}

At this point, a comment about gauge-dependencies is in order.
First of all, in the framework of large-scale dynamos, one expects scale
separation between small-scale and large-scale magnetic fields.
It is then possible to express the small-scale magnetic helicity
as a density of linkages between magnetic structures, which leads
to the manifestly gauge-invariant Gauss linking formula \citep{SB06}.
Second, the large-scale magnetic field remains in general gauge-dependent,
and there have been several examples of this \citep{BDS02,HB10}.
This would render the alternate approach of \cite{HB12} problematic,
but they argue that those gauge-dependencies result simply from a
drift in the mean vector potential and must be subtracted out.
Third, with appropriate boundary conditions, such drifts can be
eliminated, and $\bra{\AAA\cdot\BB}$ can then well reach a statistically
steady state.
If that is the case, the left-hand side of \Eq{dABm} vanishes after
time averaging, so the gauge-dependent magnetic helicity flux divergence
must balance the gauge-independent resistive term, so the former must
in fact also be gauge-independent \citep{MCCTB10,HB10}.
This argument applies even separately to the contributions from small-scale
and large-scale components; see \Eqs{dab}{dAmBm}.
This allowed \cite{DSGB13} to show for the first time that the
magnetic helicity flux divergence from the small-scale field
can become comparable to the resistive term.
Ultimately, however, one expects it of course to out-compete
the latter, but this has not yet been seen for the magnetic
Reynolds numbers accessible to date.

The issue of magnetic helicity fluxes occurs already in
the special case of a closed domain for which $\bra{\nab\cdot\meanFF}=0$,
i.e., $\oint\meanFF\cdot\dd\SSS=0$, so there is no flux in or out of
the domain, but $\meanFF$ and $\meanFFf$ can still be non-vanishing
within the domain.
This is the case especially for shear flows, where the flux term
can have a component in the cross-stream direction that is
non-uniform and can thus contribute to a finite divergence.
Simulations of \cite{HB12} have shown that such a term might be an artifact
of choosing the Coulomb gauge rather than the so-called advective gauge,
in which case such a term would vanish.
This implies that the shear-driven \cite{VC01} flux would vanish.
This term was previously thought to be chiefly responsible for
alleviating catastrophic quenching in shear flows \citep{SB04,BS04}.
It is therefore surprising that such a simple term is now removed
by a simple gauge transformation.
Clearly, more work is needed to clarify this issue further.

\subsubsection{Diffusive versus advective magnetic helicity fluxes}

To date we know of at least two types of magnetic helicity
flux that can alleviate catastrophic quenching.
One is a diffusive magnetic helicity fluxes proportional to the
negative gradient of the local value of the mean magnetic helicity density
from the small-scale fields, $\meanhf=\bra{\aaaa\cdot\bb}$, so
$\meanFFf^{\rm diff}=-\kappah\nab\meanhf$.
Another is a contribution that comes simply from advection by the mean flow
$\meanUU$, so $\meanFFf^{\rm adv}=\meanhf\meanUU$ \citep{SSSB06}.
Recent work \cite{MCCTB10}
has analyzed the contributions to the evolution equation for $\meanhf$;
see \Eq{dab}.
In the low $\Rm$ regime, the production term $2\meanEMF\cdot\meanBB$
is balanced essentially by $2\eta\mu_0\overline{\jj\cdot\bb}$.
This means that, as $\eta$ decreases, $2\meanEMF\cdot\meanBB$ must also
decrease, which leads to catastrophic quenching in that regime.
However, although the $\nab\cdot\meanFFf$ term is subdominant,
it shows a less steep $\Rm$-dependence ($\propto\Rm^{-1/2}$, as opposed
to $\Rm^{-1}$ for the $2\eta\mu_0\overline{\jj\cdot\bb}$ term),
and has therefore the potential of catching up with the other terms
to balance $2\meanEMF\cdot\meanBB$ with a less steep scaling.

Recent work using a simple model with a galactic wind has shown,
for the first time, that this may indeed be possible.
In \Fig{presults2} we show their basic result.
As it turns out, below $\Rm=100$ the $2\eta\mu_0\overline{\jj\cdot\bb}$ term
dominates over $\nab\meanFF$, but because of the different scalings
(slopes being $-1$ and $-1/2$, respectively), the $\nab\meanFF$ term is
expected to becomes dominant for larger values of $\Rm$ (about 3000).
Surprisingly, however, $\nab\meanFF$ becomes approximately constant
for $\Rm\ga100$ and $2\eta\mu_0\overline{\jj\cdot\bb}$ shows now a shallower
scaling (slope $-1/2$).
This means that the two curves would still cross at a similar value.
Our data suggest, however, that $\nab\meanFF$ may even rise slightly,
so the crossing point is now closer to $\Rm=1000$.

\subsubsection{Magnetic helicity fluxes in the exterior}

\begin{figure}[t!]\begin{center}
\includegraphics[width=.8\textwidth]{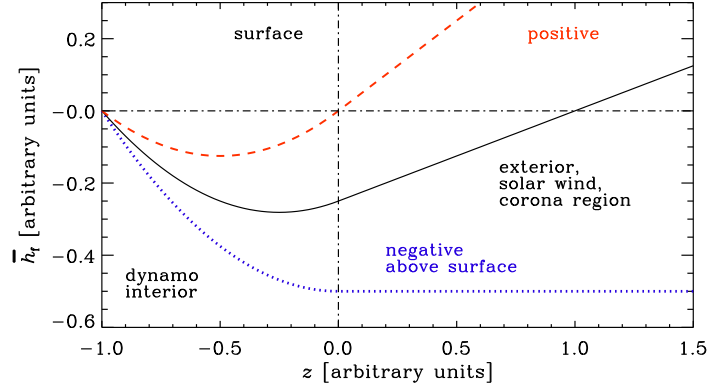}
\end{center}\caption[]{
Sketch showing possible solutions $\meanhf(z)$
with $S=\const=-1$ in $z<0$ and $S=0$ in $z>0$.
The red (dashed) and black (solid) lines show solutions for which the
magnetic helicity flux ($-\kappa_h\dd\meanhf/\dd z$)
is negative in the exterior.
The blue (dotted) line shows the case, where the magnetic helicity
flux is zero above the surface and therefore do not reverse the sign of
$\meanhf(z)$ in the exterior.
Adapted from \cite{WBM12}.
}\label{psketch}\end{figure}

Some surprising behavior has been noticed in connection with the
small-scale magnetic helicity flux in the solar wind, and it is
to be expected that such behavior also applies to galaxies.
Naively, if negative magnetic helicity from small-scale fields
is ejected from the northern hemisphere, one would expect to find
negative magnetic helicity at small scales anywhere in the exterior.
However, if a significant part of this wind is caused by a diffusive
magnetic helicity flux, this assumption might be wrong and the sign
changes such that the small-scale magnetic helicity becomes positive
some distance away from the dynamo regime.
In \Fig{psketch} we reproduce in graphical form the explanation offered
by \cite{WBM12}.

The idea is that the helicity flux is essentially diffusive in nature.
Thus, to transport positive helicity outward, we need a negative gradient,
and to transport negative helicity outward, as in the present case, we need
a positive gradient outward.
This is indeed what is shown in \Fig{psketch}.
It is then conceivable that the magnetic helicity overshoots and becomes
itself positive, which is indeed what is seen in the solar wind
\citep{BSBG11}.

\section{Observational aspects}
\label{sec:ObsAspects}

It would be an important confirmation of the nonlinear quenching theory
if one could find observational evidence for bi-helical magnetic fields.
This has not yet been possible, but new generations of radio telescopes
allow for a huge coverage of radio wavelengths $\lambda$, which could help
us determine the spatial distribution of the magnetic field using a tool
nowadays referred to as RM synthesis \citep{BB05,HBE09,FSSB11,Gie13}.
This refers to the fact that the line-of-sight integral for the complex
polarized intensity $P=Q+\ii U$ can be written as an integral over the
Faraday depth $\phi$ (which itself is an integral over $B_\|$ and, under
idealizing assumptions, proportional to the line of sight coordinate $z$)
and takes the form
\EQ
P(\lambda^2)=\int_{-\infty}^\infty F(\phi) e^{2\ii\phi\lambda^2}\,\dd\phi,
\label{Plam2}
\EN
which can be thought of as a Fourier integral for the Fourier variable
$2\lambda^2$ \citep{Bur66}.
The function $F(\phi)$ is referred to as the Faraday dispersion
function, and it would be interesting to find it by observing $P(\lambda^2)$.
The problem is of course that only positive values of $\lambda^2$
can be observed.

\begin{figure}[t!]\begin{center}
\includegraphics[width=\columnwidth]{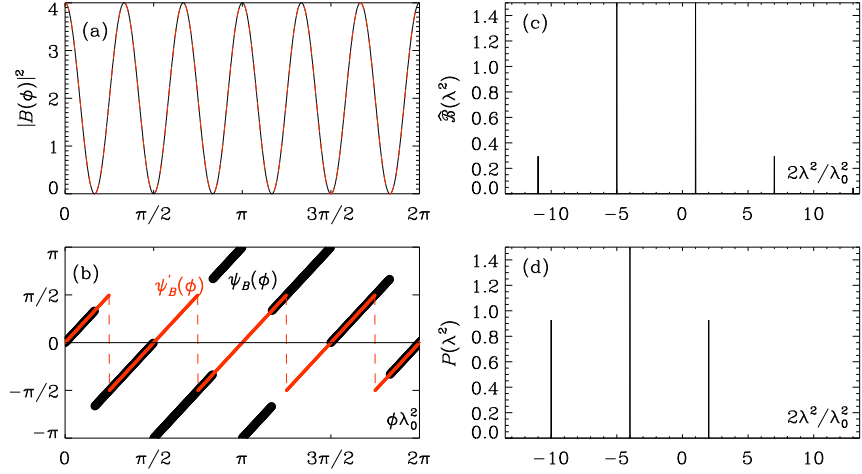}
\end{center}\caption[]{
(a) $|B|^2(\phi)$, (b) $\psi_B(\phi)$ and $\psi_B'(\phi)$,
(c) ${\cal B}(k)$, and (d) $P(k)$
for a tri-helical magnetic field with $k_2/k_1=-5$
using $\RM>0$.
In panel (b), the dashed blue lines correspond to
$\pi/2-\phi|\lambda_1^2|$ and $3\pi/2-\phi|\lambda_1^2|$
and mark the points where the phase of $\psi_B(\phi)$ jumps.
}\label{dft}\end{figure}

\begin{figure}[t!]\begin{center}
\includegraphics[width=\textwidth]{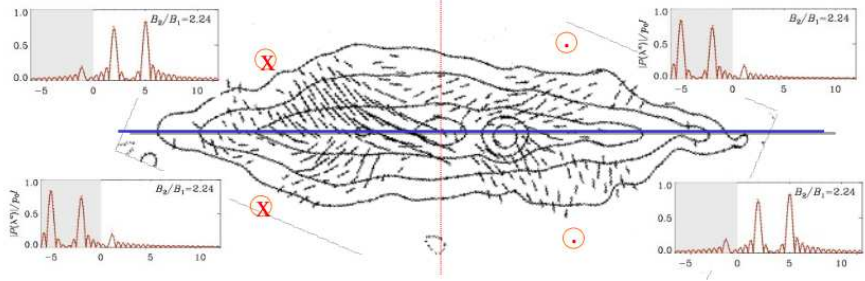}
\end{center}\caption[]{
Illustration of the four quadrants of an edge-one galaxy,
where two are expected to show signatures fields of positive
helicity and two signatures fields of negative helicity.
The $2\times2$ panels correspond to polarization maps
shown in \cite{BS14} for different signs of $\RM$ and helicity.
}\label{edge_on}\end{figure}

Most of the work in this field assumes that $P(\lambda^2)$ is Hermitian,
i.e., $P(-\lambda^2)=P(\lambda^2)^*$, where the asterisk denotes
complex conjugation.
This is however not the case for a helical magnetic field, as has recently
been pointed out \citep{BS14}.
Consider a magnetic field of Beltrami type, $\BB=(\cos kz,-\sin kz, 0)$,
write it in complex form as ${\cal B}=B_x+\ii B_y$, so that
${\cal B}(z)=B_\|\,e^{\ii\psi_B(z)}$ with $\psi_B(z)=kz$, and assume that
$\phi$ is linear in $z$ (which is the case when $n_e B_\|=\const$).
We thus obtain ${\cal B}={\cal B}(\phi(z))$.
The Faraday dispersion function is essentially given by
$F(\phi)\propto{\cal B}^2$, so its phase of now $2\psi_B$ and one loses
phase information, which is referred to as the $\pi$ ambiguity.
Inserting this into \Eq{Plam2}, one sees that most of the contribution
to the integral comes from those values of $\lambda^2$ for which
the phase is constant or ``stationary'', i.e.,
\EQ
2\ii(\psi_B+\phi\lambda^2)=\const.
\EN
Making use of the fact that for constant $n_e B_\|$ we have
$\phi=-K n_e B_\|\,z$, and thus
\EQ
\lambda^2=-k/K n_e B_\|
\EN
is the condition for the wavelength for which the integral in
\Eq{Plam2} gets its largest contribution.
Similar conditions have also been derived by \cite{Soko98} and \cite{AB11}.

The Fourier transform of such a complex field would directly reflect
the individual constituents of the magnetic field.
For example, a superposition of two helical fields yields
two corresponding peaks in the Fourier spectrum of ${\cal B}$;
see \Fig{dft}(c), where we have peaks at normalized Fourier variables
$2\lambda^2/\lambda_0^2=1$ and $-5$.
However, the quantity inferred by RM synthesis is the Faraday dispersion
function, which is related to the square of ${\cal B}$, and its Fourier
spectrum is more complicated; see \Fig{dft}(d), where we have peaks at
$2\lambda^2/\lambda_0^2=2$, $-4$, and $-10$.
Thus, the two modes combine to a new one with a Fourier variable that
is equal to the sum $1+(-5)=-4$, with side lobes separated by their
difference $1-(-5)=6$ to the left and the right.
The corresponding modulus and phase $\psi_B$ are shown in panels (a) and
(b), respectively.
Also shown is the phase $\psi_B'$, which is $\psi_B$ remapped onto
the range from $-\pi/2$ to $\pi/2$.

The magnetic fields of spiral galaxies is expected to be dominated
by a strong toroidal component.
This component might provide a reasonably uniform line-of-sight
component without reversals when viewed edge-on.
This would then provide an opportunity to detect polarization signatures
from magnetic fields with different signs of magnetic helicity in
different quadrants of the galaxy.
\FFig{edge_on} provides a sketch with a line-of-sight component of
different sign on the left of the right of the rotation axis.
On the right, $\RM$ is positive, so we can detect signatures
of the field with positive helicity.
Since the large-scale field has positive magnetic helicity in the
upper disc plane, signatures from this component can be detected
in quadrants I and III.
Conversely, since the small-scale field has positive magnetic helicity
in the lower disc plane, signatures from this component can be detected
in quadrants II and IV.

\section{Primordial magnetic field}
\label{sec:Primordial}

Primordial magnetic fields are generated in the early Universe,
either at inflation \citep{TW88} at $\la10^{-32}\s$,
the electroweak phase transition at $\sim10^{-12}\s$,
or the QCD phase transition at $\sim10^{-6}\s$;
see, for example, \cite{Vac91,Vac01} and the review by \cite{DN13}.
Such magnetic fields are basically subject to subsequent turbulent decay.
Nevertheless, the evolution of these magnetic fields is essentially
governed by the same hydromagnetic equations than those used to
describe dynamos in galaxies, for example.
The purpose of this section is to point out that there are some
important similarities between decaying turbulence in the early Universe
and (supernova-) driven turbulence in contemporary galaxies.

\begin{figure}[t]
\begin{center}
\includegraphics[width=\columnwidth]{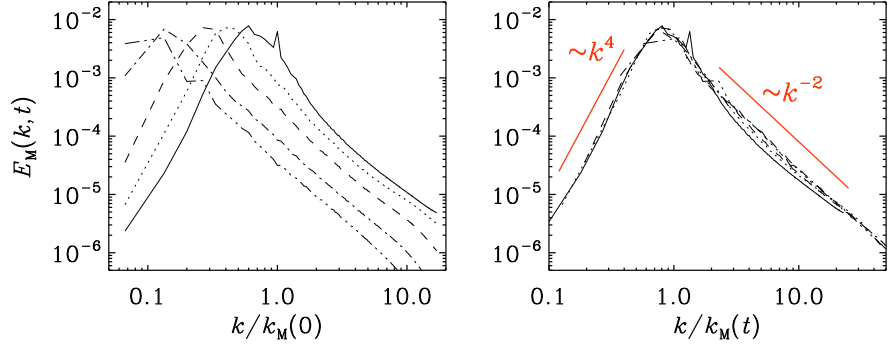}
\end{center}\caption[]{
Magnetic energy spectra at different times in the presence of magnetic
helicity with $\Pm=1$.
On the right, the abscissa is rescaled by $k_{\rm M}(t)$, which make
the spectra collapse onto each other.
The Reynolds number based on the wavenumber $k_{\rm M}$ is around 1000.
The spike at $k/k_0$ corresponds to the driving scale
prior to letting the field decay.
}\label{pkt512_helt2_short512pm1b3_noforce_both}
\end{figure}

Especially for magnetic fields generated at the electroweak phase
transition, there is the possibility that such fields are helical.
This would then lead to an inverse cascade \citep{PFL} and a transfer
of magnetic energy to progressively larger scale or smaller wavenumbers;
see \Fig{pkt512_helt2_short512pm1b3_noforce_both}, where we show
magnetic energy spectra $\EM$ versus wavenumber $k$ and compare with
the case where $k$ is divided by the integral wavenumber $\kM(t)$
defined through
\EQ
\kM^{-1}(t)=\left.\int k^{-1}\EM(k,t)\,\dd k\right/\int\EM(k,t)\,\dd k.
\EN
Simulations like those shown above are now done by several groups
\citep{CHB01,BJ04,KTBN13}.
One of the main motivations for this work is the realization that
magnetic fields generated at the time of the electroweak phase transition
would now have a length scale of just one AU, which is short compared
with the scale of galaxies.
In fact, in the radiation dominated era, the hydromagnetic equations
in an expanding universe can be rewritten in the usual form when using
conformal time and suitably rescaled quantities; see \cite{BEO96}, who
then used a magnetic helicity-conserving cascade model of hydromagnetic
turbulence to investigate the increase of the correlation length with time.

\begin{figure}[t]
\begin{center}
\includegraphics[width=\columnwidth]{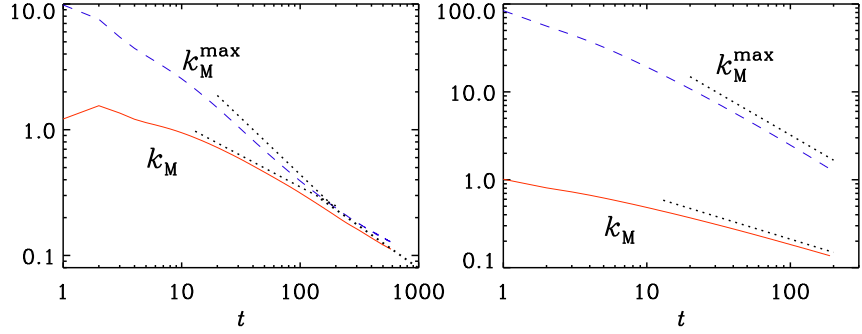}
\end{center}\caption[]{
Evolution of $\kM(t)$ (solid) and $\kM^{\max}(t)$ (dashed)
for a fractional initial helicity (left) and zero initial helicity (right).
}\label{pcomp_kft_QCD}
\end{figure}

In reality, the magnetic field will never be fully helical.
However, non-helical turbulence decays faster (like $t^{-1}$)
than helical one (like $t^{-2/3}$); see \cite{BM99}.
One can think of partially helical turbulence as a mixture of
a more rapidly decaying nonhelical component and a less rapidly
decaying helical component.
After some time, the former one will have died out and so only
the latter, helical component will survive.
In \Fig{pcomp_kft_QCD} we show the scaling of $\kM$ for weakly helical
and a nonhelical case and compare with the maximum possible value
derived from the realizability condition, i.e.
\EQ
\kM(t)\leq\kM^{\max}(t)\equiv2{\cal E}(t)/|{\cal H}(t)|,
\EN
where ${\cal E}(t)=\int\EM(k,t)\,\dd k$ and ${\cal H}(t)=\int\HM(k,t)\,\dd k$
are magnetic energy and helicity computed from the spectra.
This was originally demonstrated by \cite{TKBK12} using simulations
similar to those presented here.

The decay of helical magnetic fields is also amenable to the
mean-field treatment discussed in \Sec{MeanField}.
Helicity from large-scale fields drives helicity at small scales
via \Eq{dab} and thereby an $\alpha$ effect through \Eq{alphaKM}.
This slows down the decay \citep{YB03,KBJ11,BS13,BBS13} and may be
relevant to the survival of galactic magnetic fields, as already
mentioned in \Sec{NonAxi}.

\begin{figure}[t!]\begin{center}
\includegraphics[width=\columnwidth]{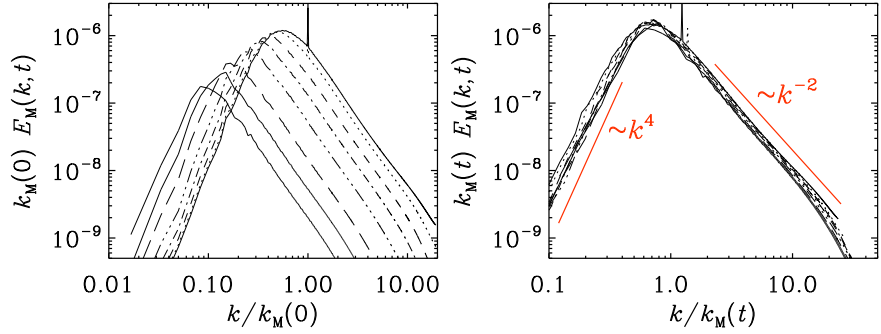}
\end{center}\caption[]{
Similar to \Fig{pkt512_helt2_short512pm1b3_noforce_both}, but for
the case without initial helicity and initial scale separation ratio
of the forcing of $k_{\min}/k_1=60$. $\Pm=1$.
On the right, the ordinate is scaled with $\kM(t)$, in addition
to the scaling of the abscissa with $1/\kM(t)$.
}\label{pkt2304_helt2_short2304pm1_kf60b_noforce_both}\end{figure}

Remarkably, simulations have shown that some type of inverse cascading
occurs also in the {\em absence} of magnetic helicity \citep{CHB01,KTBN13}.
In \Fig{pkt2304_helt2_short2304pm1_kf60b_noforce_both} we show such a
result from a simulation at a numerical resolution of $2304^3$ meshpoints.
However, the detailed reason for this inverse energy transfer still
remains to be clarified.

\section{Conclusions}

The overall significance of primordial magnetic fields is still unclear,
because contemporary magnetic fields might well have been produced by
some type of dynamo within bodies such as stars and accretion
discs within galaxies, and would then have been ejected into the rest
of the gas outside.
Whether such mechanisms would be sufficiently powerful to explain
magnetic fields even between clusters of galaxies remains to be seen.
In this connection it is noteworthy that \cite{NV10} found a lower bound
on the magnetic field strength of $3\times10^{-16}\G$ based on the
non-detection of GeV gamma-ray emission from the electromagnetic
cascade of TeV gamma rays in the intergalactic medium.
This bound is well above the even rather optimistic earlier estimated
galactic seed magnetic field strengths \citep{Ree87}.

Invoking some type of large-scale seed magnetic field seems to be the
only plausible option if one wants to explain the non-axisymmetric
magnetic fields in M81.
However, this galaxy is perhaps only one of the few where there is
still strong evidence for the existence of a non-axisymmetric magnetic
field.
In agreement with mean-field dynamo theory, most galaxies harbor
axisymmetric magnetic fields and their toroidal field is symmetric
about the midplane.

With the help of turbulent dynamo simulations over the past 20 years,
it is now clear that the conventional $\alpha\Omega$ type dynamo must
produce large-scale magnetic fields that have two different signs of
helicity, one at large scales and the opposite one at small scales.
Such magnetic fields are called bi-helical and might be detectable
through their specific signature in polarized radio emission.
These are some of the aspects that we have highlighted in the present
review about galactic dynamo simulations.
Clearly, simulations have to be conducted in close comparison with theory.
By now, simulations have reached sufficiently high magnetic Reynolds
numbers that simple theories such as first-order smoothing clearly
breaks down and some kind of asymptotic regime commences.
Given that it will not be possible to reach asymptotic scaling yet,
is must eventually be the interplay between simulations and theory
that can provide a meaningful understanding of galactic magnetism.

\begin{acknowledgement}
I am indebted to Oliver Gressel and Kandaswamy Subramanian for reading
the manuscript and providing help and useful comments.
Financial support from the European Research Council under the AstroDyn
Research Project 227952, the Swedish Research Council under the grants
621-2011-5076 and 2012-5797, as well as the Research Council of Norway
under the FRINATEK grant 231444 are gratefully acknowledged.
\end{acknowledgement}


\end{document}